\documentclass{JHEP3}

\usepackage{graphicx}
\usepackage{epsfig}

\def\br{\begin{eqnarray}}
\def\er{\end{eqnarray}}
\def\be{\begin{equation}}
\def\ee{\end{equation}}
\def\a{\alpha}

\def\<{\left\langle}
\def\>{\right\rangle}

\def\a2{\<A^{\mu}A_{\mu}\>}

\title{A dynamical gluon mass solution in a coupled system of the Schwinger-Dyson equations}
\author{A. C. Aguilar and A. A. Natale \\
Instituto de F\'{\i}sica Te\'orica, Universidade Estadual Paulista \\
Rua Pamplona 145, 01405-900, S\~ao Paulo, SP, Brazil \\ 
Email: \email{aguilar@ift.unesp.br}, 
\email{natale@ift.unesp.br}}

\abstract {We study numerically the Schwinger-Dyson equations for the coupled system of gluon and
ghost propagators in the Landau gauge and in the case of pure gauge QCD.
We show that a dynamical mass for the gluon propagator arises as
a solution while the ghost propagator develops an enhanced behavior in the infrared
regime of QCD. Simple analytical expressions are proposed for the propagators, and the
mass dependency on the $\Lambda_{QCD}$ scale and its perturbative scaling are studied. We
discuss the implications of our results for the infrared behavior of the coupling constant,
which, according to fits for the propagators infrared behavior, seems to  indicate that
$\alpha_s (q^2) \rightarrow 0$ as $q^2 \rightarrow 0$.}


\begin{document}

\section{Introduction}

Due to the property of asymptotic freedom Quantum Chromodynamics (QCD) has been
extensively tested in the regime where high energies are transferred between quarks and gluons.
On the other hand we could say that
we have only a qualitative understanding of its infrared (IR) properties. Several
 nonperturbative techniques have
been used to study the infrared region, and among these are the QCD simulations on
the lattice, which are showing strong
evidences that the gluon propagator is infrared finite \cite{mar} and that
the coupling constant may freeze at low momenta \cite{ati}.

A possible infrared finite behavior for the gluon propagator and the running coupling constant
has also been determined as a solution of the coupled system of Schwinger-Dyson equations (SDE) 
for the gluon and ghost propagators \cite{cornwall,alkofer,kondo}. Other signals for the 
freezing of the coupling constant can also be found in the
literature \cite{shirkov,ans}, and there is an accumulation of phenomenological evidences
and possible tests for this soft infrared behavior of the gluon propagator \cite{an} as well as 
for the coupling constant \cite{dok}.

In principle the Schwinger-Dyson equations provide a powerful tool to study the QCD infrared
behavior, because they comprehend an infinite tower of coupled integral equations that contain
 all the information about the theory.
In practice its intricate structure only become tractable when we make some approximations
and truncations. To illustrate the difficulty of this method we notice that only in the
nineties, Curtis and Pennington pointed out a truncation scheme, for QED, which is gauge
independent and also respect the multiplicative renormalizability \cite{CP}. At the moment
there are only indications that it is possible to do the same in QCD \cite{bloch1}, and,
unfortunately,
we have to go step by step changing and improving the approximations in order to unravel
the actual behavior, or at least to obtain rough solutions that could be compared to
other methods like QCD simulations on the lattice.

Recently we solved the SDE for the gluon propagator in the Landau gauge within the so
called Mandelstam's approximation \cite{aguilar}. This is an interesting example of how
delicate are these equations.

The gluon propagator within this approximation was first shown to behave as $1/k^4$ in the
infrared \cite{mand}, which was appraised as a clear signal of confinement. However this
result was discarded by simulations on the lattice \cite{mar}. In our previous work
\cite{aguilar} assuming a different trilinear gluon vertex and renormalization
of the final equation, as suggested by Cornwall \cite{cornwall}, and still in agreement
with the Mandelstam's approach, we obtained an infrared finite gluon propagator.
Our result showed how the approximations may change the solutions of the SDE. Of course,
although the result is very instructive it is not complete because in this approximation
the ghosts are neglected. In this work we will improve this approximation
with the inclusion of the ghosts fields.

We will solve the coupled SDE for the gluon and ghost propagators in the Landau gauge.
Fermions will not be included at this level. As in ref. \cite{aguilar} we follow
Cornwall's prescription to deal with the equation for the gluon propagator. We obtain a
numerical solution indicating the generation of a dynamical gluon mass without the
introduction of any ansatz for the solutions. The integral equation for the gluon propagator
is clearly consistent with a massive gluon polarization operator and its ultraviolet
behavior is also consistent with perturbative QCD. The distribution of our paper is
the following: In section II we build up the system of coupled equations.
In section III we discuss the angular approximation that
we perform in the integral equations, which simplifies considerably the
 amount of numerical work to solve the equations. Section IV contains a discussion
of the renormalization procedure and in section V we check
the ultraviolet behavior of the equations in order to compare it with the predictions
of perturbative QCD. In section VI
we present our numerical results and discuss its implications
for the infrared behavior of the coupling constant. Our last section contains the conclusion.

\section{The coupled SDE for the gluon and ghost propagators}

The SDE for the gluon propagator in pure gauge QCD is shown diagrammatically in figure (\ref{fullesd}).
 The gluon propagator is written in terms of itself, the full 3 and 4-point gluon vertex
$\Gamma_{\mu\nu\rho}$ and $\Gamma_{\mu\nu\rho\sigma}$, and also the full ghost propagator
and the gluon-ghost coupling. The ghost propagator depends on itself, on the gluon propagator
 and also involves the gluon-ghost coupling.


\FIGURE[t]{\centerline{\includegraphics[scale=0.6]{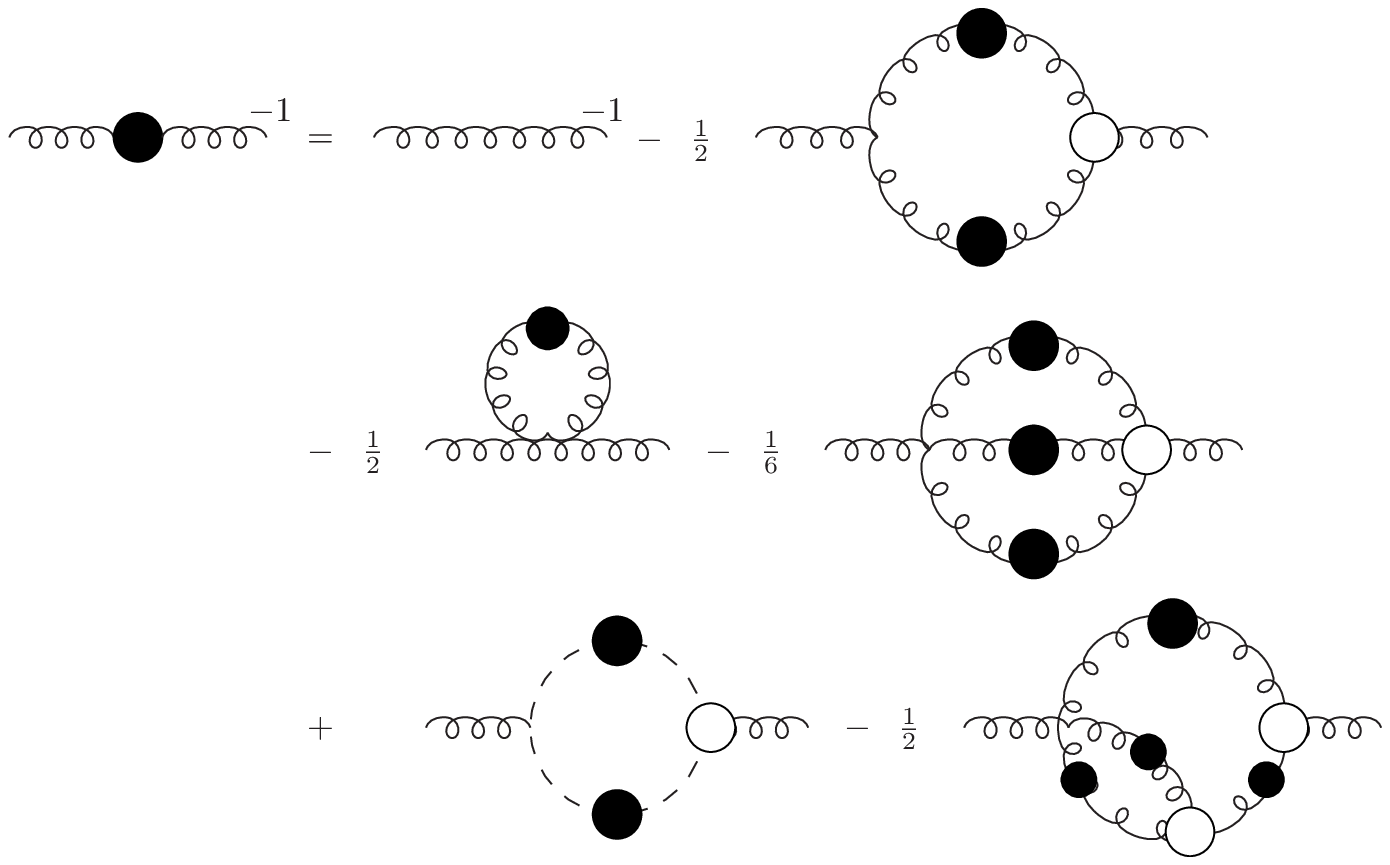}}%
\vspace{-9.5cm}
\caption{The complete Schwinger-Dyson equation for the gluon propagator without quarks. The spiral
 lines represent the gluon fields and the dashed lines the ghost fields. The black blobs
 indicate the full gluon propagators while the white ones represent the full vertices.
\label{fullesd}\label{fig1}}}


\FIGURE[t]{\centerline{\includegraphics[scale=0.6]{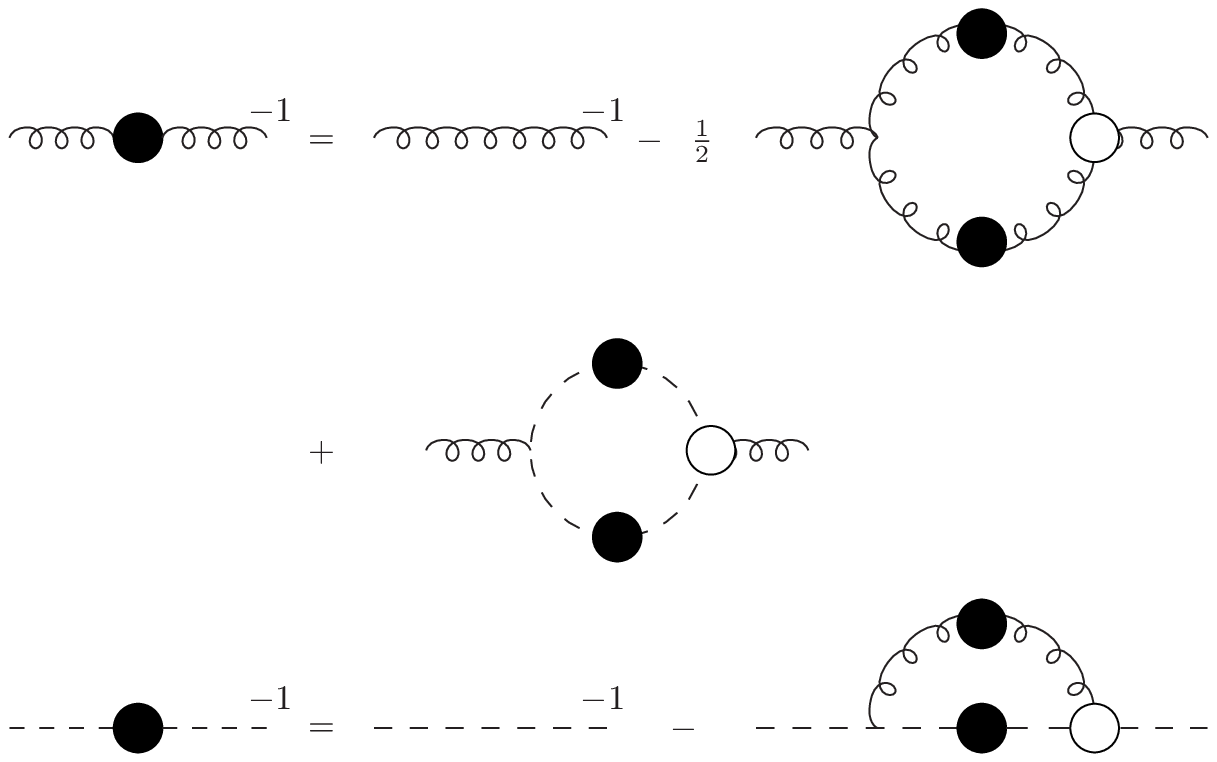}}%
\vspace{-9.5cm}
\caption{Diagrams for the coupled gluon-ghost system of Schwinger-Dyson equations.
\label{apmad}\label{fig2}}}


The  truncated renormalized SDE for the gluon propagator, in the Landau gauge and at one-loop level,
are shown in figure (\ref{apmad}) in the same approximation of ref. \cite{alkofer} and it can be
 written in the Euclidean space as

\begin{eqnarray}
D_{\mu\nu}^{-1}(k)=  Z_3D^{0}_{\mu\nu}(k) &-&
g^2C_2\tilde{Z_1}\int{\frac{d^4q}{(2\pi)^4}}iq_{\mu}D_G(p)D_G(q)G_{\nu}(q,p)\nonumber\\
 && \hspace{-1.0 cm}+
g^2C_2\frac{1}{2}Z_1\int\frac{d^4q}{(2\pi)^4}\Gamma^{0}_{\mu\rho\alpha}(k,-p,q)D_{\alpha\beta}
(q)D_{\rho\sigma}\Gamma_{\beta\sigma\nu}(-q,p,-k) ,
\label{esd_ghost}
\end{eqnarray}
where $p=k+q$, $D^{0}_{\mu\nu}(k)$ and $\Gamma^{0}_{\mu\nu\rho}(p,q,k)$
are the gluon propagator and three gluon vertex at tree level, $\Gamma _{\mu\nu\rho}(p,q,k)$
and $G_{\nu}$ are  the full three gluon and gluon-ghost vertices respectively,
and  we use the color factor $C_2=3$.  Note that all terms with four-gluon vertices were neglected.
One of these contributions is the momentum independent tadpole which can be eliminated by one
appropriate choice of the momentum projector as discussed in second paper of ref. \cite{mand}.

The full gluon propagator that enters into eq.(\ref{esd_ghost}) is expressed by
\begin{equation}\label{gluonprop}
D^{\mu\nu}(k)=\frac{\mathcal{Z}(k^2)}{k^2}\left(
\delta^{\mu\nu}- \frac{k^{\mu}k^{\nu}}{k^2}\right),
\end{equation}
where ${\mathcal{Z}(k^2)}$ is the dressing of the gluon propagator.
 When ${\mathcal{Z}(k^2)}=1$ we recover the  perturbative expression of the gluon propagator
at tree level. Therefore this function measures the transition
from the nonperturbative to the perturbative regimes as its value changes with the scale.

It will be useful to introduce here the function $D(k^2)$ that can be expressed in
terms of the dressing of the gluon propagator,
${\mathcal{Z}(k^2)}$,

\begin{equation}
D(k^2)= \frac{\mathcal{Z}(k^2)}{k^2},
\end{equation}

The full ghost propagator, $D_G(k)$, can be defined as
\begin{equation}
D_G(k) =-\frac{\mathcal{F}(k^2)}{k^2}.
\label{ghostprop}
\end{equation}
where ${\mathcal{F}(k^2)}$ is the dressing of the ghost propagator.

The renormalized SDE for the ghost propagator, shown in figure (\ref{apmad}), reads

\begin{eqnarray}
D_G^{-1}(k)= -\tilde{Z_3}k^2 +
g^2 C_2\tilde{Z_1}\int\frac{d^4q}{(2\pi)^4}ik_{\mu}D_{\mu\nu}(k-q)G_{\nu}(k,q)D_G(q).
\label{esd_ghost1}
\end{eqnarray}

Note that in the above equations, we have already introduced
$Z_3$, $\tilde{Z_3}$, $Z_1$ and $\tilde{Z_1}$ which are respectively the renormalization
constants for the gluon propagator, the ghost propagator, the three gluon and
the gluon-ghost vertices, which are needed in order to render our
coupled integral equation system finite.
Once the full three gluon and  gluon-ghost vertices are known, we have a closed coupled system
for the gluon and ghost propagators that can be solved numerically.

The construction of these vertices is based on their respective Slavnov-Taylor identities
and its form was discussed in ref. \cite{alkofer}.
Apart from a group theoretical factor ($g f^{abc}$)  we can express the full three-point
vertex function in the following way
\begin{eqnarray}
\Gamma_{\mu\nu\rho}(p,q,k)=
-A_{+}(p^2,q^2,k^2)\delta_{\mu\nu}i(p-q)_{\rho} -
A_{-}(p^2,q^2,k^2)\delta_{\mu\nu}i(p+q)_{\rho} \nonumber \\
-2 \frac{A_{-}(p^2,q^2,k^2)}{p^2-q^2}(\delta_{\mu\nu}pq
-p_{\nu}q_{\mu})i(p-q)_{\rho} + \mbox{c.p.},
\label{vertex_3g1}
\end{eqnarray}
where
\begin{equation}
A_{\pm}(p^2,q^2,k^2)= \frac{{\mathcal F}(k^2)}{2} \left(
\frac{{\mathcal F}(q^2)}{{\mathcal F}(p^2){\mathcal Z}(p^2)} \pm
\frac{{\mathcal F}(p^2)}{{\mathcal F}(q^2){\mathcal Z}(q^2)}
\right).
\end{equation}

The three gluon vertex at tree level can be recovered if we assume that
$ {\mathcal Z}(k^2)={\mathcal F}(k^2)=1 $ and in this way we
obtain $A_{-}(p^2,q^2,k^2) = 0 $ and  $A_{+}(p^2,q^2,k^2) = 1$.

The full gluon-ghost vertex can be written as \cite{alkofer}
\begin{equation}
G_{\mu}(p,q)= iq_{\mu}\frac{{\mathcal F}(k^2)}{{\mathcal F}(q^2)}
+ i p_{\mu} \left(\frac{{\mathcal F}(k^2)}{{\mathcal F}(q^2)}-1 \right).
\label{vertex_g_g}
\end{equation}

Introducing into the eq.(\ref{esd_ghost1}) the eqs.(\ref{gluonprop}),
(\ref{ghostprop}) and (\ref{vertex_g_g}), we obtain
\begin{eqnarray}
  \frac{1}{{\mathcal F}(k^2)} =
    \tilde{Z_3} - g^2C_2\tilde{Z_1}\int \frac{d^4q}{(2\pi)^4}\,
      \left( k \mathcal{P}(p) q \right) \,
      \frac{{\mathcal Z}(p^2) {\mathcal F}(q^2)}{k^2\, p^2\, q^2}
           \left( \frac{{\mathcal F}(p^2)}{{\mathcal F}(q^2)} +
 \frac{{\mathcal F}(p^2)}{{\mathcal F}(k^2)} - 1 \right) \; ,
   \label{ghost4}
\end{eqnarray}
where $p = k - q$  and
\begin{equation}
\mathcal{P}^{\mu\nu}(p) = \delta^{\mu\nu} - \frac{p^\mu p^\nu}{p^2} ,
\end{equation}
is the transversal projector.

Equations (\ref{esd_ghost}) and (\ref{ghost4}) are the coupled SDE equations for the gluon
and ghost propagators in the approximation of ref. \cite{alkofer} that we want to solve,
following the same steps performed in ref. \cite{aguilar}.

\section{The angular approximation}

In order to solve the coupled system of SDE, compound by the eqs.(\ref{esd_ghost}) and (\ref{ghost4}),
shown in the previous section, the first step is to perform the
angular integration of the coupled system of integral equations. In our previous work
\cite{aguilar}, where we considered only
the SDE for the gluon propagator and used a specific form for the three gluon
vertex,  it was simple to compute analytically
the angular integration without any
approximation. However, in the present case this is not possible anymore and  we perform an
angular approximation \cite{alkofer} as we explain in the following.

When $q^2<k^2$  we have that ${\mathcal F}(p^2)= {\mathcal F}((k-q)^2)
\rightarrow {\mathcal F}(k^2)$  and ${\mathcal Z}(p^2)
\rightarrow {\mathcal Z}(k^2) $, that preserves the correct limit when $q^2 \rightarrow 0 $ .
On the other hand when we have $q^2>k^2$
we can set ${\mathcal F}(p^2) \approx {\mathcal F}(k^2) \rightarrow {\mathcal F}(q^2)$.
According to this approximation the ghost dressing can be written as

\begin{eqnarray}
\frac{1}{{\mathcal F}(k^2)}&=&\tilde{Z_3} - \frac{9}{4}\lambda \tilde{Z_1}\left[\int_{0}^{k^2}
\frac{dq^2}{k^2}\frac{q^2}{k^2}{\mathcal Z}(k^2){\mathcal F}(k^2)
 + \int_{k^2}^{\Lambda^2}
\frac{dq^2}{q^2}{\mathcal Z}(q^2){\mathcal F}(q^2)\right],
\end{eqnarray}
where $\lambda =g^2(\mu^2)/16\pi^2$  and  the momentum integration in the SDE runs
from zero to infinity, and a ultraviolet cutoff $\Lambda$
is introduced to deal with the upper limit of the integrals.

The above equation lead us to
\begin{eqnarray}
\frac{1}{{\mathcal F}(k^2)}&=&\tilde{Z_3} - \frac{9}{4}\lambda\tilde{Z_1}
\left[\frac{1}{2}{\mathcal Z}(k^2){\mathcal F}(k^2)
 + \int_{k^2}^{\Lambda^2}
\frac{dq^2}{q^2}{\mathcal Z}(q^2){\mathcal F}(q^2)\right],
 \label{ang_aprox}
\end{eqnarray}
while the equation for the gluon propagator now reads
\begin{eqnarray}
&&\frac{1}{{\mathcal Z}(k^2)} =
     Z_3  \nonumber \\
&&+ Z_1 \lambda
     \left\{ \int_{0}^{k^2} \frac{dq^2}{k^2}
     \left( \frac{7}{2}\frac{q^4}{k^4}
     - \frac{17}{2}\frac{q^2}{k^2}
     - \frac{9}{8} \right) {\mathcal Z}(q^2) {\mathcal F}(q^2)
     +  \int_{k^2}^{\Lambda^2} \frac{dq^2}{q^2} \left(
          \frac{7}{8} \frac{k^2}{q^2} - 7 \right) {\mathcal Z}(q^2) {\mathcal F}(q^2) \right\}
 \nonumber \\
    &&  \hspace{3cm}+ \lambda \left\{ \int_{0}^{k^2}
       \frac{dq^2}{k^2}
       \frac{3}{2} \frac{q^2}{k^2} {\mathcal F}(k^2) {\mathcal F}(q^2) -
\frac{1}{3} {\mathcal F}^2(k^2)
       + \frac{1}{2} \int_{k^2}^{\Lambda^2} \frac{dq^2}{q^2}
       {\mathcal F}^2(q^2) \right\},
 \label{ghfinal}
\end{eqnarray}
where we used the projector
\begin{equation}
\mathcal{R}^{\mu\nu}(k) = \delta^{\mu\nu} -4 \frac{k^\mu k^\nu}{k^2} ,
\end{equation}

It is important to mention that to obtain eq.(\ref{ghfinal}), it was neglected one
term from the three-gluon loop when $q^2< k^2$, as discussed at length in the second and third papers
of ref.\cite{alkofer}.

The above equations are identical to the ones
obtained by the authors of ref. \cite{alkofer} in their first calculation of the coupled
SDE, and they can be viewed as a natural extension to the gluon SDE in the Mandelstam approximation,
since  the only difference between the three gluon loop contribution in the previous case and the
coupled system is that the gluon dressing function $\mathcal Z$ was replaced by the
product $\mathcal ZG$.

Our system of equations will differ from the one obtained in ref. \cite{alkofer} after the
renormalization procedure. One natural question that can arise is how much our solutions
will depend on the angular approximation introduced in this section.
Obviously, this can only be answered after improving this angular approximation, what we hope to
check
in the future, however we can say in advance that it is known
that in the case of ref. \cite{alkofer} this approximation did not introduce
a qualitative change of the solution, but it does produce a quantitative change, specially
in the infrared fixed point value of the coupling constant \cite{bloch}.

As we will discuss in detail in the section V, this angular approximation was built in order to
recover
all the parameters - the gluon and ghost anomalous dimensions and the first order coefficient of the
Callan-Symanzik $\beta(g)$ function - that describe the known perturbative behavior of the coupling
constant. However we
believe that this approximation, despite the fact that it is sucessful in describing the
perturbative
region \cite{alkofer}, can bring some numerical uncertainty in the infrared behavior, which will
be reflected
in the  ratio $m_0/\Lambda_{QCD}$,  which is a parameter,  to be
introduced in the section VI, that measures the value of the gluon propagator at zero momentum.

\section{Renormalization}

As quickly mentioned in the section III, we can link the unrenormalized propagators and
vertices with renormalized ones introducing the multiplicative renormalization
constants in the following way
\begin{eqnarray}
 D_{\mu\nu}^{nr}(q^2,\Lambda^2) &=& Z_3(\mu^2,\Lambda^2)D_{\mu\nu}(q^2,\mu^2), \nonumber \\
 D_G^{nr}(q^2,\Lambda^2) &=& \tilde{Z_3}(\mu^2,\Lambda^2)D_{G}(q^2,\mu^2),  \nonumber	\\
 g_0^{nr}(\Lambda^2) &=& \tilde{Z_g}(\mu^2,\Lambda^2)g(\mu^2),
\label{reqcd}
\end{eqnarray}
where the superscript $nr$ denotes the nonrenormalized quantity and $\mu$ is our renormalization
scale.
The renormalization constants, in the Landau gauge, are connected through the following
relations
\begin{equation}
Z_1=Z_gZ_3^{3/2} \quad \, ; \, \quad \tilde{Z_1}=Z_gZ_3^{1/2}\tilde{Z_3}.
\label{r3}
\end{equation}
Furthermore we have one more identity,  which is
satisfied only in the Landau gauge \cite{taylor}.

To determine the renormalization constants we will impose the following condition on
the ghost dressing ${\mathcal F}(\mu^2) =1$,
where $\mu$ is chosen in the high energy region, i.e. $\mu >> \Lambda_{QCD}$, 
where $\Lambda_{QCD}$ is
the QCD scale. This procedure is explained
in our previous work \cite{aguilar} and is usual when dealing with
the SDE \cite{alkofer}.

According to this we obtain
\begin{equation}\label{subt2}
{\mathcal{F}}(\mu^{2})=1,\qquad \rightarrow \qquad  \tilde{Z_{3}} =1 -
\frac{9\lambda}{4}{\mathcal {A}}_{\mathcal {F}}(\mu^{2}),
\end{equation}
where ${\mathcal {A}}_{\mathcal {F}}(x)$ is given by
\begin{eqnarray}
{\mathcal A}_{\mathcal {F}}(x)&=&\left[ \frac{1}{2}{\mathcal Z}(x){\mathcal
F}(x) + \int_{x}^{\Lambda^2} \frac{dy}{y}{\mathcal Z}(y){\mathcal
F}(y)\right],
\label{gluonb}
\end{eqnarray}
where we substituted $k^2$ by $x$ and $q^2$ by $y$.

We finally obtain the renormalized expression for the ghost SDE
\begin{equation}
{\mathcal{F}}(x)= \left[1 -\frac{9}{4}\lambda\left[ {\mathcal
A}_{\mathcal {F}}(x)-{\mathcal A}_{\mathcal {F}}(\mu^2)\right]\right]^{-1} ,
\label{gluon-ghost_renor}
\end{equation}
where ${\mathcal{F}}(x)$ in eq.(\ref{gluon-ghost_renor}) is now the renormalized ghost dressing.

In the case of the gluon propagator we can rewrite the renormalized eq.(\ref{ghfinal})
in the following compact formula
\begin{equation}
{\mathcal Z}(x)^{-1} = Z_3 + \lambda Z_1{\mathcal B}(x) + \lambda {\mathcal C}(x),
\label{compac}
\end{equation}
where
\begin{eqnarray}
{\mathcal B}(x) =  \int_0^x
\frac{dy}{x}\left(\frac{7y^2}{2x^2} -\frac{17y}{2x} -\frac{9}{8}
\right){\mathcal Z}(y){\mathcal F}(y) + \int_x^{{\Lambda}^2}
\frac{dy}{y}\left(-7
+ \frac{7x}{8y}\right){\mathcal Z}(y){\mathcal F}(y),
\end{eqnarray}
and
\begin{eqnarray}
{\mathcal C}(x) = \int_0^x \frac{dy}{x}\left(
\frac{3y}{2x} \right){\mathcal F}(y){\mathcal F}(x) +
\int_{x}^{\Lambda^2}\frac{dy}{2y}{\mathcal F}^2(y)  - \frac{1}{3}{\mathcal F}^2(x).
\end{eqnarray}

The $Z_3$ renormalization constant role is to eliminate the divergent terms of the
gluon SDE (eq.(\ref{compac})). Therefore we can add all the potentially divergent terms
imposing that
\begin{equation}
 xZ_3  + \frac{1}{2} \lambda \int_{x}^{\Lambda^2} {dy}\frac{x}{y}{\mathcal F}^2(y)
-7 \lambda\int_{x}^{\Lambda^2}dy xD(y) {\mathcal F}(y) = R + x
\label{subt}
\end{equation}
where $R$ can be determined through the gluon propagator renormalization condition
\begin{equation}
D^{-1}(\mu^2) = \mu^2  \quad \mbox{or} \quad
{\mathcal Z}(\mu^2) = 1,
\label{c34}
\end{equation}
where, again, it is worth remembering that the scale $\mu^2$ is chosen in the perturbative
region.
It is important to stress at this point that the subtractive renormalization that we have done
above is the same that has been prescribed by Cornwall many years ago \cite{cornwall}, and has
been explained in detail in ref. \cite{aguilar}. This renormalization allows a massive
solution for the gluon propagator. Moreover, as explained in \cite{cornwall} and
\cite{aguilar}, this approach
does not break the Slavnov-Taylor identity involving the gluon propagator and the trilinear
gluon vertex as long as we add to the full triple gluon vertex
massless pole terms that have been usually neglected in these equations. These terms do not
modify
the ESD but promote the consistency with the Slavnov-Taylor identities. Finally note
that we
have not discarded any term from the unrenormalized gluon SDE, but just absorbed the divergent
terms in the renormalization constant $Z_3$, and that this constant in
eq.(\ref{subt}) is proportional do $1$ plus a function of $\mu^2$ and $\Lambda^2$, which is
compatible
with the expected weak coupling expansion for this renormalization constant.

In eq.(\ref{compac}) besides the renormalization constant $Z_3$ we have some terms multiplied
by the constant $Z_1$ that comes from the trilinear vertex. We could eliminate this
constant in a rigorous way using the relations of eq.(\ref{r3}) such as
\begin{equation}
Z_1=\frac{Z_3}{\tilde{Z_3}}.
\label{r13}
\end{equation}
which follows from the identity $\tilde{Z_1}=1$.
However it was shown in ref. \cite{alkofer} that such procedure may destroy the perturbative
aspects of the solution within this approximation.
As mentioned in our previous work, ref. \cite{aguilar}, the SDE renormalization procedure
is really an intricate subject and there is not a recipe to deal with these renormalization 
constants in the nonperturbative region. In such
a situation the best that can be done is to go step by step and analyse the consequences of each
choice, and we decided in this work to choose the simplest case discussed in ref. \cite{alkofer},
$Z_1=1$. Using this relation to determine the constant $R$ defined in
eq.(\ref{subt}), we obtain the following expression
\begin{eqnarray}
R= &-&\lambda\int_{0}^{\mu^2} dy
     \left( \frac{7}{2}\frac{y^2}{\mu^4}
     - \frac{17}{2}\frac{y}{\mu^2}
     - \frac{9}{8} \right)yD(y) {\mathcal F}(y)
     -  \frac{3}{2} \lambda\int_{0}^{\mu^2} dy \frac{y}{\mu^2} {\mathcal F}(\mu^2) {\mathcal F}(y) 
\nonumber \\
&-& \lambda \int_{\mu^2}^{\Lambda^2} dy
          \frac{7}{8} \frac{\mu^4}{y^2} yD(y) {\mathcal F}(y)
 +\frac{1}{3}\lambda\mu^2{\mathcal F}^2(\mu^2).
\end{eqnarray}

We finally obtain the renormalized SDE equation for the gluon propagator
\begin{eqnarray}
D^{-1}(x)= R + x &+&\lambda\int_{0}^{x} dy
     \left( \frac{7}{2}\frac{y^2}{x^2}
     - \frac{17}{2}\frac{y}{x}
     - \frac{9}{8} \right) yD(y) {\mathcal F}(y)
     +  \frac{3}{2} \lambda\int_{0}^{x} dy \frac{y}{x} {\mathcal F}(x) {\mathcal F}(y) \nonumber \\
 &+& \lambda \int_{x}^{\Lambda^2} dy
          \frac{7}{8} \frac{x^2}{y^2} yD(y) {\mathcal F}(y)
 -\frac{1}{3}\lambda x{\mathcal F}^2(x)
\label{compac8}
\end{eqnarray}                   
where we recall that $\lambda =g^2(\mu^2)/16\pi^2$, $x=k^2$ and $y=q^2$.

It is this last equation, eq.(\ref{compac8}), together with the one for the ghost propagator,
eq.(\ref{gluon-ghost_renor}), that will be solved
numerically. Note that in eq.(\ref{compac8}) as $x \rightarrow 0$ the inverse propagator goes to
a constant value, i.e. it shows
the presence of a dynamically generated mass.

\section{Ultraviolet Behavior}

Before solving numerically the coupled SDE it is important to show that they reproduce the
QCD perturbative behavior in the high momentum region, confirming the consistency of the
renormalization procedure. In order
to do so we substitute the perturbative behavior of the gluon and ghost propagators in their
respective SDE and keep only the ultraviolet dominant terms in each equation. The solution
that comes out are

\begin{equation}
D^{-1}(x) = x\left( 1 + \frac{\gamma_0^{\prime}\alpha(\mu^2)}{4\pi}\ln\left(\frac{x}
{\mu^2}\right)\right)\,,
\label{ultrav2}
\end{equation}
and
\begin{equation}
D_G^{-1}(x) = x\left( 1 + \frac{\delta_0^{\prime}\alpha(\mu^2)}{4\pi}\ln\left(\frac{x}{\mu^2}
\right)\right).
\label{ultrav3}
\end{equation}
where $\gamma_0^{\prime}=13/2$ and  $\delta_0^{\prime}=9/4$.

As happened in the case of ref. \cite{aguilar} (see also the discussion of ref. \cite{alkofer})
the momentum behavior
of the gluon and ghost propagators is the one expected from perturbative but in addition
we recover, in this  approximation, the correct
perturbative values for the anomalous dimensions ($\gamma_0=13/2$ and  $\delta_0=9/4$ ).
Furthemore, we can also obtain the correct value for the first coefficient
of the $\beta$ function ($\beta_0^{\prime}$).

Starting from the eqs.(\ref{reqcd}) and (\ref{r3}) we can determine the running coupling
constant and $\beta_0^{\prime}$. The following relation is satisfied by the renormalized
coupling constant

\begin{eqnarray}
g(\mu^2) &=& \frac{Z_3^{3/2}(\mu^2,\Lambda^2)}{Z_1(\mu^2,\Lambda^2)}g_0^{nr}(\Lambda^2)\nonumber\\
        &=& \frac{Z_3^{1/2}(\mu^2,\Lambda^2)\tilde{Z_3}(\mu^2,\Lambda^2)}{\tilde{Z_1}(\mu^2,
\Lambda^2)}g_0^{nr}(\Lambda^2),
\end{eqnarray}
where the renormalization constants can be obtained from the gluon and ghost dressing
functions
\begin{eqnarray}
{\mathcal Z}^{nr}(x,\Lambda^2) &=& Z_3(\mu^2,\Lambda^2){\mathcal Z}(x,\mu^2) \nonumber \\
{\mathcal F}^{nr}(x,\Lambda^2) &=& \tilde{Z_3}(\mu^2,\Lambda^2){\mathcal F}(x,\mu^2),
\end{eqnarray}
leading to the following expression for the coupling constant
\begin{equation}
\alpha_s(x,\Lambda^2)=\alpha_s(\mu^2,\Lambda^2){\mathcal Z}(x,\mu^2)
{\mathcal F}^2(x,\mu^2).
\label{npg}
\end{equation}

Inverting the above definition of the coupling constant, using the perturbative definition of
the renormalized gluon and ghost functions, which are giving by eqs.(\ref{ultrav2}) and
(\ref{ultrav3}) divided by $x$; and keeping only terms $O(\alpha(\mu^2))$, we can obtain
the following expression
\begin{equation}
\frac{1}{\alpha_s(x)} = \frac{1}{\alpha_s(\mu^2)}+\frac{(\gamma_0^{\prime}+
 2\delta_0^{\prime})}{4\pi}\ln\left(\frac{x}{\mu^2}\right),
\label{as1}
\end{equation}
where $\beta_0^{\prime}= \gamma_0^{\prime}+ 2\delta_0^{\prime}= 11$, which is identical to
the perturbative value
($\beta_0 = 11$). Expressing the eq.(\ref{as1}) in a more familiar way, lead us to
\begin{equation}
\alpha_s(x) = \frac{4\pi}{\beta_0^{\prime}\ln\left(\frac{x}{\Lambda_{QCD}}\right)},
\label{alpha_pert}
\end{equation}
where $\Lambda_{QCD}$ is the usual QCD scale.

\section{Numerical Solution}

We can now turn to the numerical calculation to solve the gluon-ghost coupled-system
expressed by the
eqs.(\ref{gluon-ghost_renor}) and (\ref{compac8}). The solution comes out when we
apply an extension
of the same iterative numerical method used in ref. \cite{aguilar}.
We start defining a logarithmic grid  for the $x$ and $y$  variables  in order to perform
the integration from the deep infrared region to the high energy momenta. This grid is split
into two regions, the infrared one $[0,\mu^2]$, and the other corresponding to the perturbative 
region  $[\mu^2,\Lambda^2]$. This splitting is needed to impose
the renormalization conditions on the dressing functions at the scale $\mu^2$.

We then provide initial guesses for the functions $D(x)$ and ${\mathcal F}(x)$ and use them
to generate the coefficients of the cubic spline interpolation which will produce the
values of these functions in terms of the argument $y$. These ones are
used in the right hand side of the equation for computing the integral through Adaptive
Richardson-Romberg extrapolation.

The initial guesses are compared with the numerical results that were obtained after the
integration. The convergence criteria to stop running the numerical code is that the
difference between input and output functions must be smaller than $10^{-4}$, otherwise these
new numerical results will feed again the right-hand side of the SDE equations and re-start
all the procedure until the convergence criteria is satisfied.
Using this method we verified that our results are independent of the  starting guesses
for $D(x)$ and  ${\mathcal F}(x)$.
Our input data are the renormalization point, $\mu^2$,  and the value of the coupling
constant defined at this scale, $\alpha_s(\mu^2) = g^2(\mu^2)/4\pi$, and this will determine
the value of
$\Lambda_{QCD}$ through the eq.(\ref{alpha_pert}).
We vary both parameter $\mu^2$ and $\alpha_s(\mu^2)$ to scan how our solutions depend on
the renormalization point.

We compute the gluon propagator $D(x)$ for various sets of $\mu^2$ and $\alpha_s(\mu^2)$ within the
range $[10\,\mbox{GeV}^2,30 \,\mbox{GeV}^2]$ and $[0.22,0.30]$ respectively as shown in
 figure (\ref{f1}).  We can check through the
table(\ref{xx1}) that these values correspond to values of $\Lambda_{QCD}$ parameter from
$321\,\mbox{MeV}$ to $815\,\mbox{MeV}$. The range of momenta chosen for the scale
$\mu^2$ is consistent with a perturbative scale and
is within the window of squared momenta $10^{-6}$ to $10^6$ GeV$^2$, which is the maximum
range that our calculation can
cover without loss of precision in the infrared region.


\FIGURE[t]{\centerline{\includegraphics[scale=0.7]{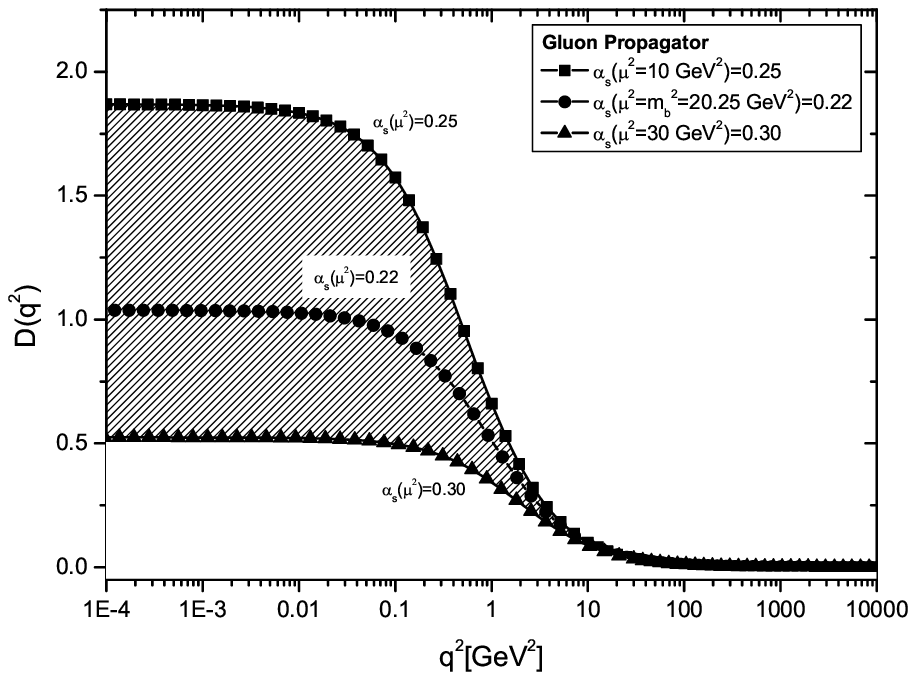}}%
\caption{Gluon propagator, $D(q^2)$, as function of momentum $q^2$ for different scales.
The line + square curve was obtained when $\alpha_s(10 \, \mbox{GeV}^2)=0.25$ which corresponds
to $\Lambda_{QCD}=  321\, \mbox{MeV}$, while in the line + triangle curve,
$\alpha_s(30 \, \mbox{GeV}^2)=0.30$, which leads to $\Lambda_{QCD}=815\, \mbox{MeV} $.
The shadowed area delimits  the curves with $\Lambda_{QCD}$ varying within the range
$ [321 \, \mbox{MeV}, 815 \, \mbox{MeV} ]$. The central curve (line + circle) was
 obtained when we fix the renormalization point, $\mu^2$, at the bottom
quark mass, $m_b^2=(4.5)^2 \, \mbox{GeV}^2$ with the central value
  $\alpha_s(m_b^2)=0.22$. 
\label{f1}\label{fig3}}}


Independently of what is the $\Lambda_{QCD}$ value we can see that all curves in figure (\ref{f1}) 
develop the same perturbative behavior, on the other hand these curves split in the infrared region
going to different values in the limit when the momentum $q^2$ goes to zero.


\FIGURE[t]{\centerline{\includegraphics[scale=0.7]{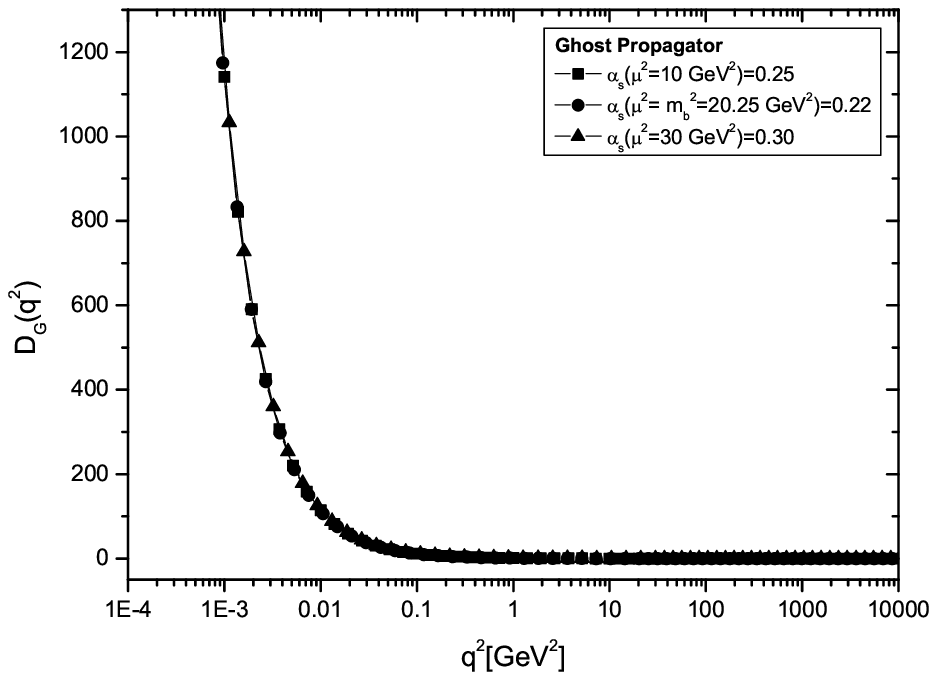}}%
\caption{Ghost propagator, $D_G(q^2)$, as function of momentum $q^2$  for the same scales
 shown in the figure (\ref{f2}). The line + square was
obtained when $\alpha_s(10 \, \mbox{GeV}^2)=0.25$ which corresponds  $\Lambda_{QCD}=  321\,
\mbox{MeV}$, while that  in the
line + triangle curve, $\alpha_s(30 \, \mbox{GeV}^2)=0.30$, which leads to
$\Lambda_{QCD}=815\, \mbox{MeV} $. The line + circle was
obtained when we fix the renormalization point, $\mu^2$, at bottom quark mass,
$m_b^2=(4.5)^2 \, \mbox{GeV}^2$ with the central value
$\alpha_s(m_b^2)=0.22$. Such curves are obtained varying the $\Lambda_{QCD}$
parameter within the range $[321 \, \mbox{MeV}, 815 \,\mbox{MeV} ]$.
\label{f4}\label{fig4}}}


For the same set of parameters we have also plotted, in the figure (\ref{f4}), the behavior of the 
ghost propagator as a function of the momentum $q^2$. As we can see, for the ghosts, the infrared
behavior has a slighter dependence on the renormalization point value than the gluon propagator.
In the overall, all curves develop the same behavior from the deep infrared to the high energy 
region.

In the figures showing the behavior of the gluon and ghost propagators,
figures (\ref{f1}) and (\ref{f4}), the most representative curve is the
one where the renormalization point, $\mu^2$, is set at the bottom quark mass,
$m_b^2=(4.5)^2 \, \mbox{GeV}^2$ with
$\alpha_s(m_b^2)=0.22$. In this case we obtain a value for $\Lambda_{QCD}=335\, \mbox{MeV} $.
We are going to concentrate only in this curve to analyse the perturbative and also non-perturbative
regimes, however it is important to keep in mind that it does not really matter what curve we choose 
to study the high energy regime, since, as mentioned before, all of them have the same perturbative
behavior.

For the input discussed above, we plot in  figure (\ref{f2}) the gluon propagator, $D(q^2)$, together
with its ultraviolet behavior described by eq.(\ref{ultrav2}) while, in  figure (\ref{f6}),
we compare the ghost propagator with its ultraviolet behavior expressed by eq.(\ref{ultrav3}).
It is interesting to note that, in this latter case, the behavior of the nonperturbative ghost 
propagator is not
so much different from its perturbative behavior, since the major difference
starts happening only for $q^2$ values less than $2\times 10^{-3}  \, \mbox{GeV}^2$.


\FIGURE[t]{\centerline{\includegraphics[scale=0.7]{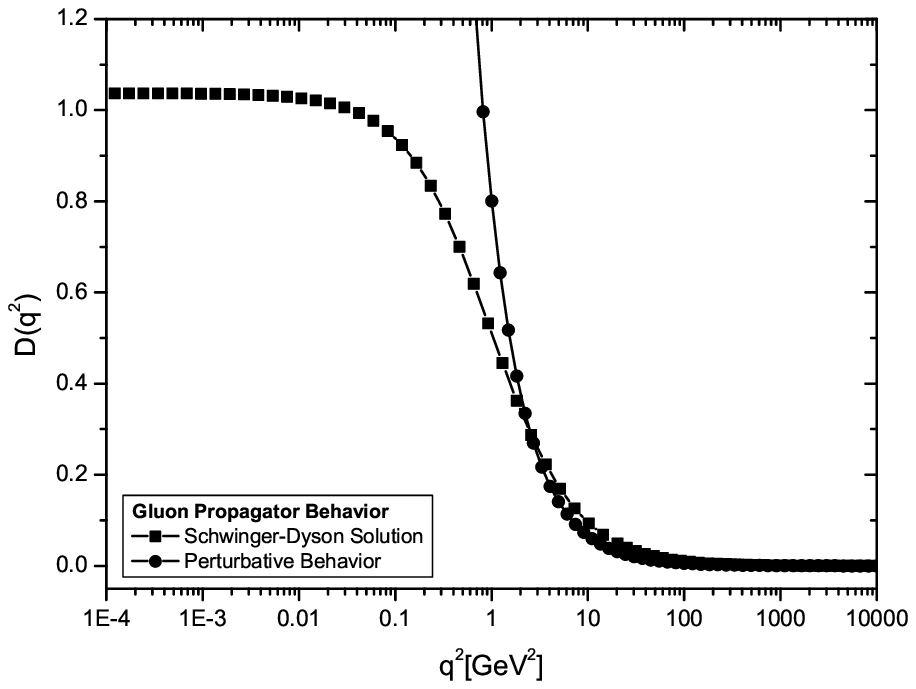}}%
\caption{The behavior of  the gluon propagator, $D(q^2)$, obtained through
the numerical solution of the Schwinger-Dyson equations,
when \mbox{${\alpha_s(\mu^2)=0.22}$} at $\mu^2=(m_{\mbox{b}})^2= (4.5)^2\, \mbox{GeV}^2$,
 together with its ultraviolet behavior
given by eq.(\ref{ultrav2}) where $\beta_0=11$.
\label{f2}\label{fig5}}}


Despite the fact that the ghost fields are important to warrant the gluon transversality in
the perturbative region and also to recover the first order beta function coefficient and anomalous
dimension of the propagators, the above result suggests that neglecting this field, as happens in
the Mandelstam approximation, based on an extrapolation from its known small contribution in the
perturbative region can be a reasonable approximation.


\FIGURE[t]{\centerline{\includegraphics[scale=0.7]{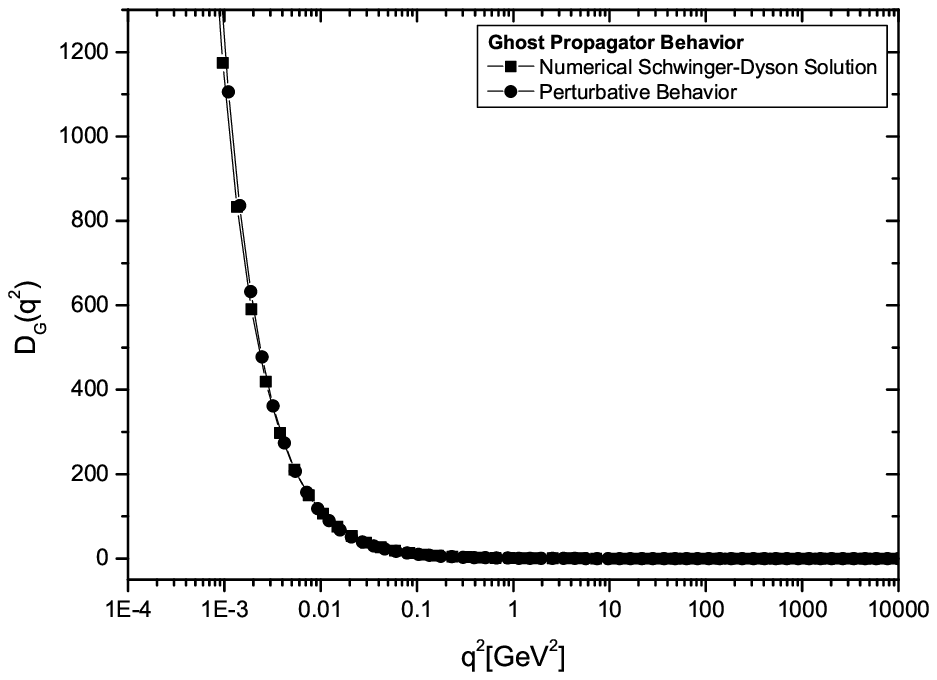}}%
\caption{The behavior of the ghost propagator, $D_G(q^2)$, obtained through the numerical
solution of the Schwinger-Dyson equation,
when \mbox{${\alpha_s(\mu^2)=0.22}$} at $\mu^2=(m_{\mbox{b}})^2= (4.5)^2\, \mbox{GeV}^2$,
together with its ultraviolet behavior
given by eq.(\ref{ultrav3}) where $\beta_0=11$.
\label{f6}\label{fig6}}}

In order to obtain a value for the gluon propagator near $q^2 = 0$, we propose a very simple fit
for our numerical data, which in Euclidean space can be written as
\begin{equation}
D(q^2)= \frac{1}{q^2 + {\mathcal M}^2(q^2)},
\label{prop}
\end{equation}
and where the dynamical gluon mass ${\mathcal M}^2(q^2)$ is described by
\begin{equation}
{\mathcal M}^2(q^2)= \frac{m_0^4}{q^2+ m_0^2}.
\label{ope}
\end{equation}

The reason for this fit was discussed in our previous work \cite{aguilar}.
In figure (\ref{f3}) the numerical solution for the gluon propagator, determined with the
renormalization point fixed at the bottom quark mass, is quite
well adjusted with the above fit for $m_0^2= 0.99\, \mbox{GeV}^2$.
In table (\ref{xx1}) the values of $m_0^2$,
utilized in the eq.(\ref{ope}), are shown for each value of the renormalization scale and
coupling constant. The value of $m_0^2$ itself is not important because it is linked to the
 scale $\Lambda_{QCD}$ through the running
of the coupling constant. What really matters is the analysis of the ratio
$m_0 / \Lambda_{QCD}$ which ranges
from $1.72$ to $2.97$ and in agreement with previous determinations for this
ratio \cite{cornwall,an}. It is clear
that different fits will give slightly different values for the gluon mass,
but other choices, as long as they
respect the correct asymptotic limits, do not differ appreciably from the values we
quoted above. Furthermore
the ratio is also relatively stable against variations of the renormalization point.


\FIGURE[t]{\centerline{\includegraphics[scale=0.7]{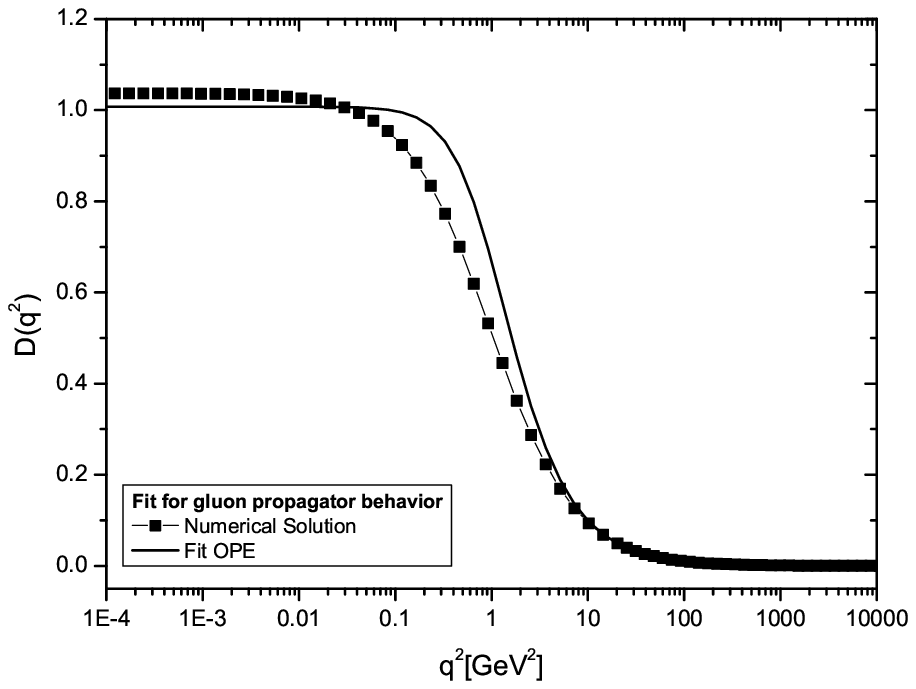}}%
\caption{Numerical solution for the gluon
propagator, $D(q^2)$, versus momentum $p^2$ for
$\alpha_s(m^2_{b})=0.22$. We compare this numerical solution
with the fit given by eq.(\ref{prop}), where  $m_0^2= 0.99\, \mbox{GeV}^2$.}
\label{f3}\label{fig7}}

We can also discuss the behavior of the coupling constant in the infrared. In order to do
so we start remembering that the running coupling constant is given by eq.(\ref{npg}), where
the gluon and ghost dressing functions are utilized as input.  Note that if we assume the
fit given by eq.(\ref{prop}) for the gluon propagator and roughly approximate the ghost dressing
 function
by $1$, because in the full region of momenta its behavior is almost the perturbative
one as can be seen in figure (\ref{f6}), we clearly obtain a vanishing coupling in
the deep infrared region.
Since the major difference between the nonperturbative and the perturbative behavior of the
ghost propagator starts happening for  $q^2$ values less than $2 \times 10^{-3}$, instead of the
rough approximation ${\mathcal{F}}(x) \approx 1$ in the low momenta region we introduce the
following fits: ${\mathcal{F}}(x) = (x/m_0^2)^\eta$ for the ghost dressing function, and
${\mathcal{Z}(x)} = (x/(x+m_0^2))^\delta$ for the gluon dressing function, in the region delimited
by $q^2 < 0.1 \, \mbox{GeV}^2$. These exponents provide additional freedom to our fits,
and we are able to fit the data with $\eta=-0.04$ and $\delta=0.98$. These figures confirm the
same vanishing behavior of the coupling constant in the deep infrared region previously examined, in
the context of lattice QCD simulation, by the authors of Ref.\cite{orsay-roma}.

\TABLE[t]{
\centerline{
\begin{tabular}{ccccc}
   \hline
  \hline
   $\alpha(\mu^2)\quad $ & $\mu ^2$ & $\Lambda_{QCD}\quad$ & $m_0^2$ & $m_0/\Lambda_{QCD}$\\
    &&$(\beta=11)\quad$&& \\
   \hline
   $0.22\quad$ & $20.25\, \mbox{GeV}^2$  & $335 \,\mbox{MeV}\quad$ & $0.99 \, \mbox{GeV}^2$ & $2.97 \,\qquad$\\
   $0.25\quad$ & $10\, \mbox{GeV}^2$  & $321 \,\mbox{MeV}\quad$ & $0.55\, \mbox{GeV}^2$ & $2.31 \,\qquad$\\
   $0.25\quad$ & $20\, \mbox{GeV}^2$  & $455 \,\mbox{MeV}\quad$ &$1.10\,\mbox{GeV}^2$ & $2.31 \,\qquad$ \\
   $0.25\quad$ &$30\, \mbox{GeV}^2$&  $557 \,\mbox{MeV}\quad$ & $1.65\, \mbox{GeV}^2$ &$2.31 \,\qquad$ \\
   $0.30\quad$ & $10\, \mbox{GeV}^2$   & $471 \,\mbox{MeV}\quad$ &$0.65\, \mbox{GeV}^2$ & $1.72 \,\qquad$ \\
   $0.30\quad$ & $20\, \mbox{GeV}^2$   & $666 \,\mbox{MeV}\quad$ &$1.31\, \mbox{GeV}^2$ & $1.72 \,\qquad$ \\
   $0.30\quad$ & $30\, \mbox{GeV}^2$  & $815 \,\mbox{MeV}\quad$ &$1.96\, \mbox{GeV}^2$ & $1.72\,\qquad$\\
    \hline
   \hline
 \end{tabular}}
\caption{Values of the renormalization point, $\mu^2$, and coupling constant,
$\alpha(\mu^2)$, used as input data in the eqs.(\ref{gluon-ghost_renor}) and (\ref{compac8}).
These values fix automatically, through eq.(\ref{alpha_pert}),
the $\Lambda_{QCD}$ scale.  In the third column, we have the values of $\Lambda_{QCD}$
computed  with  our perturbative value of  $\beta_0=11$. The values for the ratio 
$m_0/\Lambda_{QCD}$ are also shown in the last column where the $m_0^2$ values are defined
 through eq.(\ref{ope}).\label{xx1}\label{tab1}}}
%

\section{Conclusions}

We solved the coupled system of Schwinger-Dyson equations for the gluon and ghost propagators in
the Landau gauge in the case of pure gauge QCD. The new ingredient in our approach is that we use a
renormalization prescription formulated by Cornwall many years ago which allows for a dynamically
generated mass. As explained in a previous work this solution is compatible with the Slavnov-Taylor
identities when a new piece containing massless poles is added to the triple gluon full vertex.
The renormalization constant has the form $Z_3 = 1 + f(\mu^2,\Lambda^2)$ which is consistent with
the weak coupling expansion of this constant. The ratio between the dynamical gluon mass and
$\Lambda_{QCD}$ is also consistent with previous determinations.

To obtain the SDE solutions we do not need any ansatz for the asymptotic equation and we are able
to solve them numerically in a quite large range of momenta. The numerical solutions are quite stable
and we could, as we did in ref. \cite{aguilar}, compute an effective potential for composite operators
to show that we also obtain a reasonable value for the gluon condensate, although this calculation is
a little bit redundant because the gluon propagator do not differ appreciably from the one that we
obtained in the previous work. It seems that in the present case the inclusion of ghosts induces 
only
a minor numerical modification of the previous result. We present simple fits for the
gluon and ghost propagator and discuss the infrared behavior of the running coupling constant.
These fits allow us to study the behavior of the running coupling in the deep infrared region, and 
indicate that the running coupling goes to zero when $q^2 \rightarrow 0$. 

There are many points that still must be improved in the present approach. We need to compute the
equations without the angular approximation. A different approximation, other than $Z_1 = 1$ must
be tested. The inclusion of fermions and the tadpole diagram are among the many other
modifications that must be considered in the future.

\section{Acknowledgments}

We benefited from discussions with A. Cucchieri and G. Krein and we would also like to thank A.
Colato for his numerical hints. This research was supported by the Conselho Nacional de
Desenvolvimento Cient\'{\i}fico e Tecnol\'{o}gico (CNPq) (AAN) and by
Funda\c{c}\~ao de Amparo \`{a} Pesquisa do Estado de S\~{a}o
Paulo (FAPESP) (ACA).


\begin {thebibliography}{99}

\bibitem{mar} P.~Marenzoni, G.~Martinelli, N.~Stella, e M.~Testa,
Phys.\ Lett.\ {\bf B318} (1993) 511; C. Alexandrou, Ph. de Forcrand and E. Follana, 
Phys. Rev. {\bf D65} (2002) 114508; {\bf D65} (2002) 117502; F. D. R. Bonnet {\it et al.},
Phys. Rev. {\bf D64} (2001) 034501; {\bf D62} (2000) 051501; D. B.
 Leinweber {\it et al.} (UKQCD Collaboration), Phys. Rev. {\bf D58} (1998) 031501;
C. Bernard, C. Parrinello, and A.
Soni, Phys. Rev. {\bf D49} (1994) 1585; see also the most recent simulation of
P. O. Bowman {\it et al.}, hep-lat/0402032 and the
references therein.


\bibitem{ati} A recent study about the freezing of the coupling constant in a SU(2) gauge
theory with references to QCD simulations can be found in J. C. R. Bloch, A. Cucchieri, K. Langfeld
and T. Mendes, hep-lat/0312036; Nucl. Phys.
Proc. Suppl. {\bf 119} (2003) 736.


\bibitem{cornwall} J. M. Cornwall, Phys. Rev. {\bf D26} (1982) 1453; J. M. Cornwall and
J. Papavassiliou, Phys. Rev. {\bf D40} (1989) 3474; {\bf D44} (1991) 1285.


\bibitem{alkofer} R. Alkofer and L. von Smekal, Phys. Rept. {\bf 353} (2001) 281;
L. von Smekal, A. Hauck and R. Alkofer, Ann.
Phys. {\bf 267} (1998) 1; A. Hauck, L. von Smekal and R. Alkofer, 
Comput. Phys. Commun. {\bf 112} (1998) 166; L. vonSmekal, A. Hauck and R. Alkofer, Phys. Rev. Lett. {\bf 79} (1997) 3591.

\bibitem{kondo} K.-I. Kondo, hep-th/0303251.

\bibitem{shirkov} D. V. Shirkov and I. L. Solovtsov, Phys. Rev. Lett. {\bf 79} (1997) 1209;
 D. V. Shirkov, Eur. Phys. J. {\bf C22} (2001) 331;
 D.V. Shirkov, Theor. Math. Phys. {\bf 132} (2002) 1309; 
 V. N. Gribov, Nucl. Phys. {\bf B139} (1978) 1;
see also the interesting review by Y. L. Dokshitzer and D. E. Kharzeev, hep-ph/0404216
where the Gribov's results about infrared finite gluon propagator and coupling constant 
are revisited.

\bibitem{ans}A. C. Aguilar, A. A. Natale  and P. S. Rodrigues da Silva, 
Phys.\ Rev.\ Lett.\  {\bf 90} (2003)
 152001.

\bibitem{an} F. Halzen, G. Krein and A. A. Natale, Phys. Rev. {\bf D47} (1993) 295;
M. B. Gay Ducati, F. Halzen
and A. A. Natale, Phys. Rev. {\bf D48} (1993) 2324;
A. Mihara and A. A. Natale, Phys. Lett. {\bf B482} (2000) 378; 
A. C. Aguilar, A. Mihara and A. A. Natale, Int. J. Mod. Phys. {\bf 19} (2004) 249.


\bibitem{dok} A. C. Mattingly and P. M. Stevenson, Phys. Rev. {\bf D49} (1994) 437;
Y. L. Dokshitzer,
G. Marchesini and B. R. Webber, Nucl. Phys. {\bf B469} (1996) 93; Y. L. Dokshitzer,
in {\sl Proc. 29th Int.
Conf. on High Energy Physics} (ICHEP 98), Vancouver, Canada, 1998, High energy physics,
 Vol.1, p. 305,
hep-ph/9812252; S. J. Brodsky, hep-ph/0310289; A. C. Aguilar, A. Mihara and A. A. Natale, Phys. Rev.
{\bf D65} (2002) 054011.


\bibitem{CP} D. C. Curtis and M. R. Pennington, Phys. Rev. {\bf D42} (1990) 4165.

\bibitem{bloch1} See  J. C. R. Bloch, Few. Body Syst. {\bf 33} (2003) 111 and references therein.

\bibitem{aguilar} A. C. Aguilar and A. A. Natale, hep-ph/0405024

\bibitem{mand} S. Mandelstam, Phys. Rev. {\bf D20} (1979) 3223;
N. Brown and M. R. Pennington, Phys. Rev. {\bf D38} (1988) 2266; Phys. Rev. {\bf D39} (1989) 2723.

\bibitem{bloch} D. Atkinson and J. C. R. Bloch, Mod. Phys. Lett. {\bf A13} (1998) 1055.

\bibitem{taylor} J. C. Taylor, Nucl. Phys. {\bf B33} (1971) 436.

\bibitem{orsay-roma} P. Boucaud, J. P. Leroy, J. Micheli, O. Pene
and C. Roiesnel, JHEP {\bf 10} (1998) 17; P. Boucaud {\it et al.} JHEP {\bf 201} (2002) 46; 
Nucl. Phys. Proc. Suppl. {\bf 106} (2002) 266. 


\end{thebibliography}
\end{document}